\documentstyle[prl,aps,epsf]{revtex} \def\narrowtext{} \tighten \twocolumn
\input epsf.sty
\begin{document}
\draft

\title{Superconducting Gap Anisotropy and Quasiparticle Interactions: a Doping
Dependent ARPES Study}
\author{
        J. Mesot,$^{1,2}$
        M. R. Norman,$^1$
        H. Ding,$^3$
        M. Randeria,$^4$
        J. C. Campuzano,$^{1,2}$
        A. Paramekanti,$^4$
        H. M. Fretwell,$^2$
        A. Kaminski,$^2$
        T. Takeuchi,$^5$
        T. Yokoya,$^6$
        T. Sato,$^6$
        T. Takahashi,$^6$
        T. Mochiku,$^7$ and K. Kadowaki$^8$
       }
\address{
         (1) Materials Sciences Division, Argonne National Laboratory,
             Argonne, IL 60439\\
         (2) Department of Physics, University of Illinois at Chicago,
             Chicago, IL 60607\\
         (3) Department of Physics, Boston College, Chestnut Hill, 
             MA  02467\\
         (4) Tata Institute of Fundamental Research, Mumbai 400005, 
             India\\
         (5) Department of Crystalline Materials Science, Nagoya 
             University, Nagoya 464-01, Japan\\
         (6) Department of Physics, Tohoku University, 980-8578 Sendai, 
             Japan\\
         (7) National Research Institute for Metals, Sengen, Tsukuba,
             Ibaraki 305, Japan\\
         (8) Institute of Materials Science, University of Tsukuba,
             Ibaraki 305, Japan\\
         }
\address{%
\begin{minipage}[t]{6.0in}
\begin{abstract}
Comparing ARPES measurements on Bi2212 with penetration depth data, we show
that a description of the nodal excitations of the d-wave superconducting state
in terms of non-interacting quasiparticles is inadequate, and we estimate the
magnitude and doping dependence of the Landau interaction parameter which
renormalizes the linear T contribution to the superfluid density. Furthermore,
although consistent with d-wave symmetry, the gap with underdoping cannot be
fit by the simple cos$k_x$-cos$k_y$ form, which suggests an increasing
importance of long range interactions as the insulator is approached.
\typeout{polish abstract}
\end{abstract}
\end{minipage}}

\maketitle
\narrowtext

There is little doubt about the fundamental importance of many-body interactions
in high temperature cuprate superconductors \cite{ANDERSON}.
Quantifying these interactions is difficult in the normal state of these
materials, given the lack of well-defined single-particle excitations as 
revealed by various experiments.  On the other hand, well-defined 
quasiparticle excitations do exist in the superconducting state, 
and it is believed that a description of the low temperature state in terms of 
superfluid Fermi liquid theory is appropriate.  
In Fermi liquid theory, the quasiparticles are characterized by
a renormalized Fermi velocity $v_F$, and their residual interactions
described by Landau parameters,
which manifest themselves through a renormalization of
various response functions relative to that given by a non-interacting theory.
For example, in the cuprates, the Fermi velocity $v_F$ has been determined by
angle resolved photoemission (ARPES) studies in
Bi$_2$Sr$_2$CaCu$_2$O$_{8+\delta}$ (Bi2212)\cite{OLSON} to be
renormalized by a factor of two to three over that given by band theory.
The strong renormalization of the superfluid density $\rho_s(0)$ has
also been known for some time, where one sees a scaling with the number of
doped holes: the Uemura relation\cite{UEMURA}.

In this paper we examine an issue which is at the heart of the nature of 
quasiparticles in the superconducting state of the cuprates, that is, 
whether the slope of the superfluid density at low temperatures, 
$d\rho_s/dT$, is affected by interactions or not, and what the relation of its 
renormalization is to that of $\rho_s(0)$, questions of considerable debate in
the recent literature\cite{LEE,MILLIS}. The importance of $\rho_s(T)$ to an
understanding of cuprate superconductivity derives from the early observation 
of a linear $T$ suppresion of $\rho_s(T)$ \cite{HARDY}, since this is explained 
most naturally by the thermal excitations of quasiparticles near the nodes of a 
d-wave superconducting gap. Related to this is the interesting question of 
whether the gap around the node scales with $T_c$, as has been suggested 
from a recent analysis of magnetic penetration depth data\cite{PANA1}.

To address these issues we use the unique capability of ARPES to 
directly measure the Fermi wavevector $k_F$, velocity $v_F$, and the 
superconducting gap anisotropy near the node, from which we can estimate 
the slope of $\rho_s(T)$ assuming non-interacting quasiparticles. Comparing 
this with the actual value obtained by penetration depth experiments 
leads to a direct estimate of the renormalization due to quasiparticle
interactions.  This is done by exploiting the relation\cite{MILLIS}
\begin{equation}
\left|{d\rho_s \over {dT}}(T=0)\right|\propto
\left|{d \over dT}\left({1 \over \lambda^2}\right)\right|
= A\beta^2\frac{v_Fk_F}{v_{\Delta}}.
\label{1}
\end{equation}
where $\lambda$ is the penetration depth,
and $A$ is a doping-independent constant:
$A = 4\ln2\alpha k_Bn/cd$ with $\alpha$
the fine structure constant, $k_B$ the Boltzmann constant, $c$ the 
speed of light, and $n$ the number of $CuO_{2}$ layers (4 for Bi2212)
per c-axis lattice constant $d$ (30.9 \AA\ for Bi2212).
ARPES is used to determine the three parameters at the node:
the Fermi velocity $v_F$, the Fermi wavevector $k_F$,
and the slope of the superconducting gap
$v_{\Delta}=1/2|d\Delta/d\phi|(\phi=\pi/4)$, 
where $\phi$ is the Fermi surface angle.  
The latter is normalized such that $v_{\Delta}=\Delta_{\rm max}$ for the
simple d-wave gap $\Delta(\phi)=\Delta_{\rm max}\cos(2\phi)$. 

The only unknown in Eq.~1 is the renormalization factor
$\beta$ due to quasiparticle interactions;
in the isotropic Fermi liquid theory
$\beta = 1 + F_{1s}/2$, where
$F_{1s}$ is the $l=1$ spin symmetric Landau parameter, and quantifies
the backflow of the medium around the quasiparticles\cite{NOZIERES}.
By comparing ARPES and penetration depth data, we estimate $\beta$
and its doping dependence. In particular, different assumptions in 
the recent literature \cite{LEE,MILLIS,PANA1} about the doping 
dependence of $v_{\Delta}$ has led to different conclusions regarding 
the value and doping dependence of $\beta$ in Eq.~1. 

Our main results are as follows.
(1) We determine the doping dependence of the gap anisotropy from
ARPES.  Although consistent with a node on the Fermi surface along the zone
diagonal ($\phi = \pi/4$) for all doping levels, the shape of the gap
changes with underdoping: while its maximum value 
increases\cite{HARRIS,SNS97,TUNNEL},
we find the new result that the gap becomes flatter
near the nodes, i.e. $v_{\Delta}$ decreases. 
(2) Using our data on the doping dependence of $v_{\Delta}$, we 
exploit Eq.~1 and use available values of the penetration depth $\lambda(T)$ 
\cite{HARDY,BONN,WALDRAM,WALDMANN,PANA2} to estimate the renormalization
factor $\beta$.  We find that 
$\beta$ is considerably smaller than unity and decreases with underdoping,
in contrast to previous suggestions in the literature \cite{LEE,MILLIS,PANA1}.
(3) Our results on the doping dependence of the gap anisotropy and its
relation to penetration depth data provide important evidence that the
strength of both the pairing interaction and the quasiparticle interactions
increase with reduced doping.

The ARPES experiments were performed at the Synchrotron Radiation Center,
Wisconsin, using both a high-resolution 4-meter normal incidence and plane
grating monochromators, with a resolving power of $10^{4}$ at $10^{11}$
photons/sec. We used 22 eV photons, with a 17 meV (FWHM) energy
resolution, and a momentum window of radius 0.045$\pi$
(in units of $1/a$ where $a$ is the Cu-Cu separation).
The high quality single crystal samples were float-zone grown,
with doping changed by varying the oxygen partial pressure during annealing.
All samples show sharp x-ray diffraction rocking curves and flat
surfaces after cleaving as determined from specular laser reflection.
We label the samples by their doping (UD for underdoped, OD for overdoped)
and onset $T_c$.

Fig.~1 shows ARPES data at $T$=15 K for an UD75K sample at
different ${\bf k}$-points along the Fermi surface.
$k_{F}$ was carefully chosen using the criterion that the leading
edge of the spectrum has minimum binding energy with the
steepest slope, when compared with other spectra along a cut perpendicular
to the Fermi surface, as discussed earlier\cite{DING97}.
The zero of binding energy ($E_F$)
was determined from the spectra (not shown) of a polycrystalline Pt reference
in electrical contact with the Bi2212, recorded at regular
intervals to ensure accurate determination of the Fermi energy,
$E_{F}$. From the shift of spectral weight away from $E_F$, one
clearly sees an anisotropic gap, which is
maximal near the $(\pi,0)$ point ($\phi = 0$) and zero near
the $(\pi,\pi)$ direction ($\phi = 45^\circ$)\cite{SHEN93,RAPID96}.

\begin{figure}
\epsfxsize=3.0in
\epsfbox{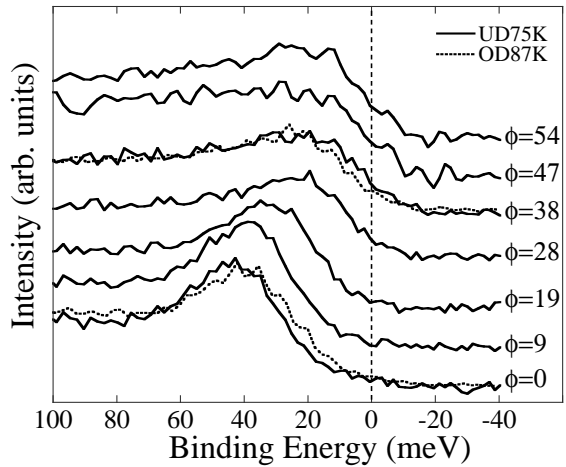}
\vspace{0.5cm}
\caption{
Spectra of an UD75K Bi2212 sample (solid line) in the vicinity of
$E_{F}$ taken at T=15K, each labeled by the Fermi surface angle
$\phi$.  For two angles we also
plot spectra from an OD87K sample (dotted line).
}
\label{fig1}
\end{figure}

\begin{figure}
\epsfxsize=3.4in
\epsfbox{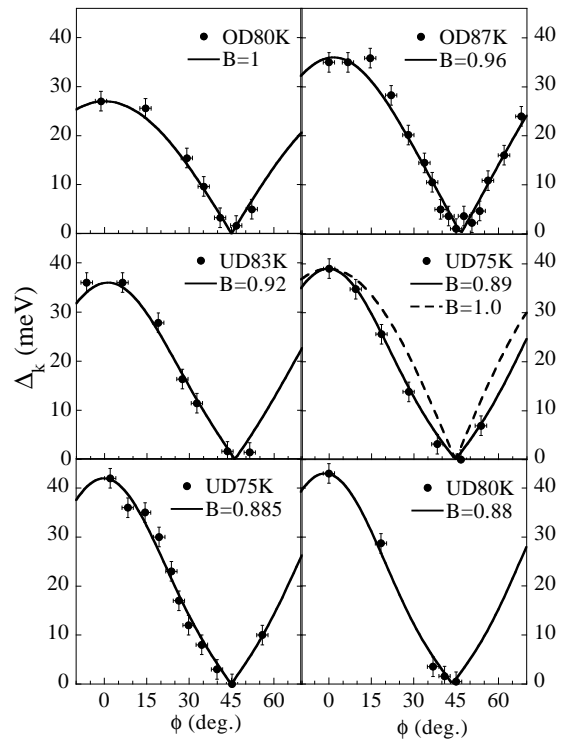}
\caption{
Values of the superconducting gap as a function of the Fermi surface
angle $\phi$
obtained for a series of Bi2212 samples with varying doping.
Note two different UD75K samples were measured, and
the UD83K sample has a larger doping due to aging\protect\cite{DING97}.
The solid lines represent the best fit using the gap function:
$\Delta_{k}=\Delta_{\rm max}[B\cos(2\phi)+(1-B)\cos(6\phi)]$
as explained in the text.
The dashed line in the panel of an UD75K sample represents the gap
function with B=1.
}
\label{fig2}
\end{figure}

For comparison we also plot (dashed line) in Fig.~1
ARPES spectra from an OD87K sample at two points on the Fermi surface.
(For more OD data see Ref.~\onlinecite{RAPID96}.) We immediately
see that the UD sample has a larger maximum gap ($\phi=0$)
than the OD one, but it has a smaller gap at the corresponding point
($\phi = 38$ degrees) near the node.  Thus the raw data directly give
evidence for an interesting change in gap anisotropy with doping.

To quantitatively estimate the gap, we have
modeled the low temperature data by a simple BCS spectral function,
taking into account the measured dispersion and the known energy
and momentum resolutions. Details of this analysis,
and error estimates, have been described earlier in
the context of OD samples \cite{DING1,RAPID96}.
The resulting angular dependence of the gap is plotted in Fig.~2 for 
six samples.

\begin{figure}
\epsfxsize=3.35in
\epsfbox{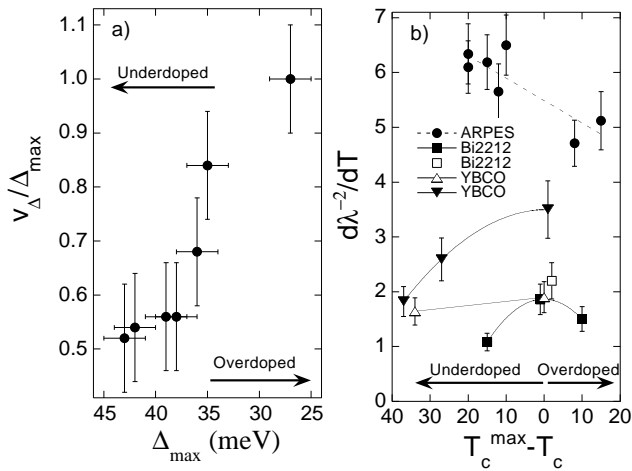}
\vspace{0.5cm}
\caption{
(a) Normalized slope of the gap at the node 
($v_{\Delta}/\Delta_{\rm max}$) vs gap maximum $\Delta_{\rm max}$.
Note the clear drop from unity as one enters the underdoped regime.
(b) Slope of the superfluid density (in units of $10^{-9}\AA^{-2}K^{-1}$) 
vs $T_{c}^{max}-T_{c}$ estimated from ARPES measurements based on
non-interacting quasiparticles in Bi2212 (filled 
circles) compared with direct penetration depth measurements in YBCO 
(open-\protect\cite{BONN} and filled-\protect\cite{PANA2} triangles,
$T_c^{max}$=92K) and Bi2212 (open-\protect\cite{WALDRAM} and 
filled-\protect\cite{WALDMANN} squares, $T_c^{max}$=95K).
The error bars for the latter values were $\pm 15\%$ based on $\pm 5\%$ 
error bars for $\lambda(0)$ \protect\cite{WALDMANN}.  The lines are
guides to the eye.
}
\label{fig3}
\end{figure}

To further quantify this change in anisotropy,
we have used the following expression to fit the gap:
$\Delta_{\bf k}=\Delta_{\rm max}[B\cos(2\phi)+(1-B)\cos(6\phi)]$
with $0\leq B \leq 1$, where $B$ is determined for each data set.
Note that $\cos(6\phi)$ is the next harmonic consistent with $d$-wave 
symmetry.  We find that while the overdoped data sets are consistent with  
$B\approx 1$,
the parameter $B$ decreases significantly in the underdoped regime.
To emphasize the significance of $B < 1$,
we plot in the panel of an UD75K sample of Fig.~2 a dashed curve with 
$B = 1$ along with the best fit curve for that sample.  From these fits,
one easily determines the value $v_{\Delta}$ discussed earlier in the
context of Eq.~1.
In Fig.~3a, we plot $v_{\Delta} / \Delta_{\rm max}$ for seven samples
(the six analysed above
plus an UD85K sample from Ref.~\onlinecite{NAT98}).  One can clearly
see from this figure the trend that underdoping leads to an increase
in the maximum gap together with a decrease in the gap slope at the node.

Several questions need to be addressed before proceeding further. 
First, could the flattening at the node be, in fact,
evidence for a ``Fermi arc'' (a line of gapless excitations),
especially since such arcs are seen above $T_{c}$
in the underdoped materials \cite{NAT98}?
Given the error bars on gap estimates in Fig.~2, it is impossible to
rule out arcs in all the samples. Nevertheless, it is clear that
there are samples (especially OD87K, UD80K and UD75K) where there
is clear evidence in favor of a point node rather than an arc
at low temperatures. Furthermore, it is very important to note
that a linear $T$ dependence of $\rho_s(T)$ at low
temperature, for all doping levels, in clean samples gives independent
evidence for point nodes \cite{HARDY,BONN,PANA2}.

Second, is the change in gap anisotropy intrinsic, or related to impurity
scattering?  We can eliminate the latter explanation on two grounds.
The maximum gap {\it increases} as the
doping is reduced, opposite to what would be expected from pair breaking due to
impurities.  Also, impurity scattering is expected to lead to a
characteristic ``tail" to the leading edge \cite{FN}, for which there
is no evidence in the observed spectra (see Fig.~1).

We suggest that the change in the gap function with underdoping is
related to an increase in the range of the pairing interaction:
the $\cos(6\phi)$ term in the Fermi surface harmonics
can be shown to be closely related to the tight binding function
$\cos(2k_x) - \cos(2k_y)$, which represents next nearest neighbours 
interaction, just as $\cos(2\phi)$ is closely related to the near neighbor
interaction $\cos(k_x)-\cos(k_y)$. On very general grounds, the
increasing importance of the $\cos(6\phi)$ term with underdoping
could arise from a decrease in screening as one approaches the 
insulator.  Similar effects also arise in specific models.
In models of spin-fluctuation mediated
d-wave pairing, an increase in the antiferromagnetic correlation
length with underdoping leads to a more sharply peaked
pairing interaction in ${\bf k}$-space, causing a flattening of the gap around
the node as we find here.  In interlayer tunneling models,
one also expects changes in the shape of the gap which might be 
correlated with doping \cite{PWA}.  

We note that the ratio of the dispersion
normal to the Fermi surface ($v_F$) to that along the Fermi
surface ($v_{\Delta}$) is quite large, 20 in the overdoped
case, and becomes even {\it larger} as the doping decreases, in
contrast to the undoped insulator which exhibits an isotropic
dispersion about the $(\pi/2,\pi/2)$ points\cite{WELLS}.  This implies that
the electronic dispersion in the superconductor in this region of the
zone may not be as closely related to the insulator as has been 
recently suggested\cite{RONNING}.

We now return to Eq.~1.
It is known from previous ARPES measurements that the 
band dispersion along $(0,0)-(\pi,\pi)$ is rather strong
and doping independent \cite{MARSHALL} with an estimated 
$v_{F}=2.5 \times 10^{7}$ cm/sec \cite{DING1}.
It is also known that $k_F$ along this direction is
0.737 $\AA^{-1}$ and relatively doping independent \cite{DING97}.
Using these inputs, together with the strongly doping dependent
$v_{\Delta}$, we can estimate the slope 
$\left|d\lambda^{-2}/dT\right|$ in the case of non-interacting 
quasiparticles $(\beta=1)$. As shown in Fig.~3b (filled circles) we 
find the resulting slope
is of order $6 \times 10^{-9} \AA^{-2}K^{-1}$ and is reduced by approximately
30\% in going from UD75K to OD87K.

Fig.~3b also shows the values of $\left|d\lambda^{-2}/dT\right|$
obtained from London penetration depth
measurements \cite{HARDY,BONN,WALDRAM,WALDMANN,PANA2}. 
Although, there is considerable variation in the measured values
of $\lambda(0)$ and low temperature ${d\lambda}/{dT}$
from one group to another, probably due to the use of different
techniques, we find evidence for the following trend: the slope $d\rho_{s}/dT$
{\it decreases} with underdoping. For YBCO this effect is weak in the
UBC data \cite{BONN}, but much stronger in the Cambridge data \cite{PANA2}.
The limited data available for Bi2212 are consistent with this
trend \cite{WALDMANN}.
The striking feature is that, in all cases, this trend in
$d\rho_{s}/dT$ is exactly the opposite of that deduced from a theory
with non-interacting quasiparticles $(\beta=1)$ using ARPES input.
That is, from Fig.~3b, it is clear that the renormalization factor $\beta$
is considerably smaller than unity and doping dependent, a conclusion
different from that inferred earlier\cite{LEE,MILLIS,PANA1}.
To get an estimate of the
doping dependence of $\beta$, we use the Bi2212 values of
Ref.~\onlinecite{WALDMANN} for OD85K and UD80K samples in comparison to
our own values on OD87K and UD80K, obtaining a $\beta^2$ of 0.32 and 0.17,
respectively.  This is roughly consistent with a $\beta$ which varies as
$x$, the number of doped holes, which would be the expected result from
the $x$ scaling of $\rho_s(0)$\cite{LEE,MILLIS}.  On the other hand, as
noted earlier\cite{MILLIS}, a weaker doping dependence of $\beta$ seems
to be implied by the UBC data.  Given the difficulties of measuring the
superconducting gap in YBCO by ARPES, this points to the need for further
penetration depth experiments on Bi2212 samples so that a more detailed
comparison to ARPES data can be made.

In conclusion, we find that the gap anisotropy of Bi2212 changes strongly
as a function of doping, implying an increase in the range of the pairing
interaction with underdoping.  Moreover, a comparison of our data to
penetration depth measurements indicates that the slope of the superfluid
density is renormalized by a doping dependent factor, implying that a
non-interacting picture of quasiparticle excitations around the nodes of
the d-wave order parameter is inappropriate.  This has obvious implications
for other low temperature measurements in the high temperature cuprate
superconductors, such as specific heat, NMR, and microwave and thermal
conductivity, which are usually quantified by theories which do not take into
account these renormalizations.

This work was supported by the the U. S. Dept. of Energy,
Basic Energy Sciences, under contract W-31-109-ENG-38, the National
Science Foundation DMR 9624048, and
DMR 91-20000 through the Science and Technology Center for
Superconductivity, and the CREST of JST.  JM is supported by the Swiss
National Science Foundation, and MR by the
Swarnajayanti fellowship of the Indian DST.


\begin{references}

\bibitem{ANDERSON}
P. W. Anderson, {\it The Theory of Superconductivity in the High $T_c$
Cuprates} (Princeton Univ. Pr., Princeton, 1997).

\bibitem{OLSON}
C. G. Olson {\it et al.}, Phys. Rev. B {\bf 42}, 381 (1990).

\bibitem{UEMURA}
Y. J. Uemura {\it et al.}, Phys. Rev. Lett. {\bf 62}, 2317 (1989).

\bibitem{LEE}
P. A. Lee and X.-G. Wen, Phys. Rev. Lett. {\bf 78}, 4111 (1997)
and {\bf 80}, 2193 (1998).

\bibitem{MILLIS}
A. J. Millis, S. M. Girvin, L. B. Ioffe, and A. I. Larkin,
J. Phys. Chem. Solids {\bf 59}, 1742 (1998).

\bibitem{HARDY}
W. N. Hardy {\it et al.}, Phys. Rev. Lett. {\bf 70}, 3999 (1993).

\bibitem{PANA1}
C. Panagopoulos and T. Xiang, Phys. Rev. Lett. {\bf 81}, 2336 (1998).

\bibitem{NOZIERES}
P. Nozieres, {\it Theory of Interacting Fermi Systems} (Addison-Wesley, Reading,
1964), p. 10; A. J. Leggett, Phys. Rev. {\bf 140}, A1869 (1965).

\bibitem{HARRIS}
J.M. Harris {\it et al.}, Phys. Rev. B {\bf 54}, R15665 (1996).

\bibitem{SNS97}
H. Ding {\it et al.}, J. Phys. Chem. Solids {\bf 59}, 1888 (1998).

\bibitem{TUNNEL}
Ch. Renner {\it et al.}, Phys. Rev. Lett. {\bf 80}, 149 (1998);
N. Miyakawa {\it et al.}, Phys. Rev. Lett. {\bf 80}, 157 (1998).

\bibitem{BONN}
D. A. Bonn {\it et al.}, Czech. J. Phys. {\bf 46}, 3195 (1996).

\bibitem{WALDRAM}
S.-F. Lee {\it et al.}, Phys. Rev. Lett. {\bf 77}, 735 (1996).
See also T. Jacobs {\it et al.}, Phys. Rev. Lett. {\bf 75}, 4516 (1995).

\bibitem{WALDMANN}
O. Waldmann {\it et al.}, Phys. Rev. B {\bf 53}, 11825 (1996). This 
measurement is restricted to $T > 17$ K.

\bibitem{PANA2}
C. Panagopoulos, J. R. Cooper, and T. Xiang, Phys. Rev. B {\bf 57}, 
13422 (1998).

\bibitem{DING97}
H. Ding {\it et al.}, Phys. Rev. Lett. {\bf 78}, 2628 (1997).

\bibitem{SHEN93}
Z.-X. Shen {\it et al.}, Phys. Rev. Lett. {\bf 70}, 1553 (1993).

\bibitem{RAPID96}
H. Ding {\it et al.}, Phys. Rev. B {\bf 54}, R9678 (1996).

\bibitem{DING1}
H. Ding {\it et al.}, Phys. Rev. Lett. {\bf 74}, 2784 (1995).

\bibitem{NAT98}
M. R. Norman {\it et al.}, Nature {\bf 392}, 157 (1998).

\bibitem{FN}
R. Fehrenbacher and M. R. Norman, Phys. Rev. B {\bf 50}, 3495 (1994).

\bibitem{PWA}
S. Chakravarty, A. Sudbo, P. W. Anderson, and S. Strong, Science {\bf 
261}, 337 (1993).

\bibitem{WELLS}
B. O. Wells {\it et al.}, Phys. Rev. Lett. {\bf 74}, 964 (1995).

\bibitem{RONNING}
F. Ronning {\it et al.}, Science {\bf 282}, 2067 (1998).

\bibitem{MARSHALL}
D.S. Marshall {\it et al.}, Phys. Rev. Lett. {\bf 76}, 4841 (1996).

\end{references}
\end{document}